\documentclass[twocolumn]{aastex63}

\usepackage{amsmath}
\usepackage{natbib}
\usepackage{color,soul}
\usepackage{url}
\usepackage{lineno}

\definecolor{LightCyan}{rgb}{0.88,1,1}

\newcommand{\lenstool}{{\tt{Lenstool}}}

\newcommand{\TEMPLATES}{TEMPLATES}
\newcommand{\templateslong}{Targeting Extremely Magnified Panchromatic Lensed Arcs and their Extended Star formation}

\newcommand{\HSTlong}{{\it Hubble Space Telescope}}

\newcommand{\zspec}{z_{spec}}

\newcommand{\lya}{Ly$\alpha$}

\newcommand{\HeI}{\hbox{{\rm He}\kern 0.1em{\sc i}}}
\newcommand{\MgI}{\hbox{{\rm Mg}\kern 0.1em{\sc i}}}
\newcommand{\MgII}{\hbox{{\rm Mg}\kern 0.1em{\sc ii}}}
\newcommand{\FeII}{\hbox{{\rm Fe}\kern 0.1em{\sc ii}}}
\newcommand{\kms}{km s$^{-1}$}

\newcommand{\clustername}{SDSS\,J1226}
\newcommand{\clusternameN}{SDSS\,J1226$+$2152}
\newcommand{\clusternameS}{SDSS\,J1226$+$2149}
\newcommand{\arcnamelong}{SGAS\,J122651.3$+$215220}
\newcommand{\zcluster}{0.4358}
\newcommand{\zarc}{2.9233}

\begin{document}

\title{HST-Based Lens Model of SDSS\,J1226$+$2152, in Preparation for JWST-ERS TEMPLATES\footnote{Based on observations made with the NASA/ESA {\it Hubble Space Telescope}, obtained at the Space Telescope Science Institute, which is operated by the Association of Universities for Research in Astronomy, Inc., under NASA contract NAS 5-26555. These observations are associated with programs GO-12166, GO-12368, GO-15378}}

\correspondingauthor{Keren Sharon}
\email{kerens@umich.edu}

\author[0000-0002-7559-0864]{Keren Sharon}
\affiliation{Department of Astronomy, University of Michigan, 1085 S. University Ave, Ann Arbor, MI 48109, USA}

\author[0000-0002-8261-9098]{Catherine Cerny}
\affiliation{Centre for Extragalactic Astronomy, Durham University, South Road, Durham DH1 3LE, UK}
\affiliation{Institute for Computational Cosmology, Durham University, South Road, Durham DH1 3LE, UK}

\author[0000-0002-7627-6551]{Jane R. Rigby}
\affiliation{Observational Cosmology Lab, Code 665, NASA Goddard Space Flight Center, 8800 Greenbelt Rd., Greenbelt, MD 20771, USA}

\author[0000-0001-5097-6755]{Michael K. Florian}
\affiliation{Steward Observatory, University of Arizona, 933 North Cherry Ave., Tucson, AZ 85721, USA}

\author[0000-0003-1074-4807]{Matthew B. Bayliss}
\affiliation{Department of Physics, University of Cincinnati, Cincinnati, OH 45221, USA}

\author[0000-0003-2200-5606]{H{\aa}kon Dahle}
\affiliation{Institute of Theoretical Astrophysics, University of  Oslo,  P. O. Box 1029, Blindern, N-0315 Oslo, Norway }

\author[0000-0003-1370-5010]{Michael D. Gladders}
\affiliation{Department of Astronomy and Astrophysics, University of Chicago, 5640 South Ellis Avenue, Chicago, IL 60637, USA}
\affiliation{Kavli Institute for Cosmological Physics, University of Chicago, 5640 South Ellis Avenue, Chicago, IL 60637, USA}

\author[0000-0003-3266-2001]{Guillaume Mahler}
\affiliation{Centre for Extragalactic Astronomy, Durham University, South Road, Durham DH1 3LE, UK}
\affiliation{Institute for Computational Cosmology, Durham University, South Road, Durham DH1 3LE, UK}
\author[0000-0003-4470-1696]{Kate A. Napier}
\affiliation{Department of Astronomy, University of Michigan, 1085 S. University Ave, Ann Arbor, MI 48109, USA}

\begin{abstract}

The strongly lensed $z=2.9233$ Lyman break galaxy,
SGAS\,J122651.3$+$215220, lensed by a cluster at $z=0.4358$, was first
targeted by JWST as part of the JWST-ERS program TEMPLATES: Targeting Extremely Magnified Panchromatic Lensed Arcs and their Extended Star formation. Aiming to combine the exquisite capabilities of JWST with the extreme magnification provided by strong gravitational lensing, these observations will peer into galaxies at cosmic noon and probe the building blocks of star formation. Here, we present a Hubble Space Telescope-based strong lensing analysis, lens model, source-plane interpretation, and the lensing outputs needed to analyze the JWST observations in the context of the source's intrinsic properties. The lens model outputs are made publicly available to the community through the Mikulski Archive for Space Telescopes (MAST) Portal.

\end{abstract}


\section{Introduction} \label{sec:intro}
Strong gravitational lensing has become an invaluable tool, routinely used in combination with high resolution space-based, adaptive-optics, or interferometry enhanced imaging and spectroscopy to obtain unprecedented spatial resolutions of galaxies at high redshift \citep[see][for a review]{Kneib2011}. This combination allows us to reveal and resolve the internal structure of galaxies at cosmic noon, when the Universe formed most of its stars, and study the physical conditions within star-forming regions at scales that cannot be otherwise probed at these redshifts. 
The JWST-ERS program \TEMPLATES: \templateslong\ (program number 1355, PI: Rigby) will observe four strongly lensed galaxies at the peak of galaxy assembly, $z=1-4$, to spatially resolve key diagnostics of star formation and extinction, on source plane scales of $\sim100$ pc. The selected targets have extensive ground-based and space-based data, and are well characterized, providing significant leverage for interpreting the new observations.

The first science target to have been observed by JWST was one of these four \TEMPLATES\ program galaxies: \arcnamelong, a $z=$\zarc\ Lyman break galaxy \citep{Koester2010}, strongly lensed by the foreground cluster \clusternameN, which is one of several sub-clusters in a complex large-scale structure at $z=\zcluster$. 
The lensed galaxy was discovered by \citet{Koester2010} as part of the Sloan Giant Arcs Survey \citep[SGAS, PI: Gladders; ][]{Hennawi2008,Bayliss2011,Sharon2020}, as a bright ($r=20.6$ mag) $u$-band dropout. Their follow-up observations determined the spectroscopic redshift of the main lensed galaxy in the field, as well as a nearby companion galaxy at the same redshift, other lensed sources, and several cluster galaxies \citep{Koester2010,Bayliss2011}. 
The foreground cluster complex contains three cluster-scale structures within $2\farcm5$ in projection and a few hundred \kms\ in velocity \citep{Bayliss2014,oguri2012}. Two of the cores are strong lenses: the north cluster, \clusternameN, and the south cluster, \clusternameS\ \citep[also a MACS cluster;][]{Ebeling2001}, both lens several background sources into giant arcs. \citet{Bayliss2014} obtained spectroscopic redshifts of hundreds of galaxies in the field, and measured a velocity dispersion of $\sigma_v=870\pm60$ \kms\ from $98$ cluster members within a projected radius of $1.5$ Mpc. They further identified several groups along the line of sight, which likely enhance the lensing cross section of this structure. 

The high strong lensing (SL) magnification acting on the source galaxy \arcnamelong\ allowed in-depth ground-based spectroscopic studies of the star formation diagnostics of the source galaxy \citep{Wuyts2012,Malhotra2017,rigby2018,Chisholm2019,Solimano2021} and its environment \citep{Solimano2022}. The clumpy source and its proximity to the critical curve were used to explore the properties of dark matter in the foreground lens \citep{Dai2020}. The bright arcs were also used as backlight to study the circumgalactic medium around a foreground galaxy, from absorption lines in the spectra of the bright arcs \citep{Tejos2021,Mortensen2021}.

To fully exploit the magnification enhancement of cosmic telescopes requires a detailed understanding of the gravitational lensing properties of the lens itself, to translate the observed measurements to their un-lensed intrinsic properties. A robust measurement of the lensing magnification, and its uncertainties, is essential for converting the observed luminosity, star formation rate, and stellar mass to their source-plane values. The lensing analysis interprets and translates between observed image plane and unobserved source plane geometry, required for measuring the physical sizes of star forming clumps, global morphology, and the physical separation between source components. 

In this paper, we present the details of a strong lensing analysis and lens model of the complex structure lensing \arcnamelong, based on archival HST imaging, to accompany the pre-launch high level science products delivery of the JWST-ERS \TEMPLATES\ program. The lens modeling outputs and the fully reduced HST mosaics are made available to the community through the Mikulski Archive for Space Telescopes (MAST) Portal via \dataset[10.17909/zqax-2y86]{\doi{10.17909/zqax-2y86}}\footnote{\url{https://archive.stsci.edu/hlsp/templates/}}. 

We assume a flat cosmology with $\Omega_{\Lambda} = 0.7$, $\Omega_{m}=0.3$, and $H_0 = 70$ \kms\ Mpc$^{-1}$. In this cosmology, $1''=5.6541$ kpc at the cluster redshift, $z=$\zcluster, and $1''=7.7616$ kpc at the source redshift, $z=$\zarc. Magnitudes are reported in the AB system unless otherwise stated.

\begin{figure*}
\epsscale{1.1}
\plotone{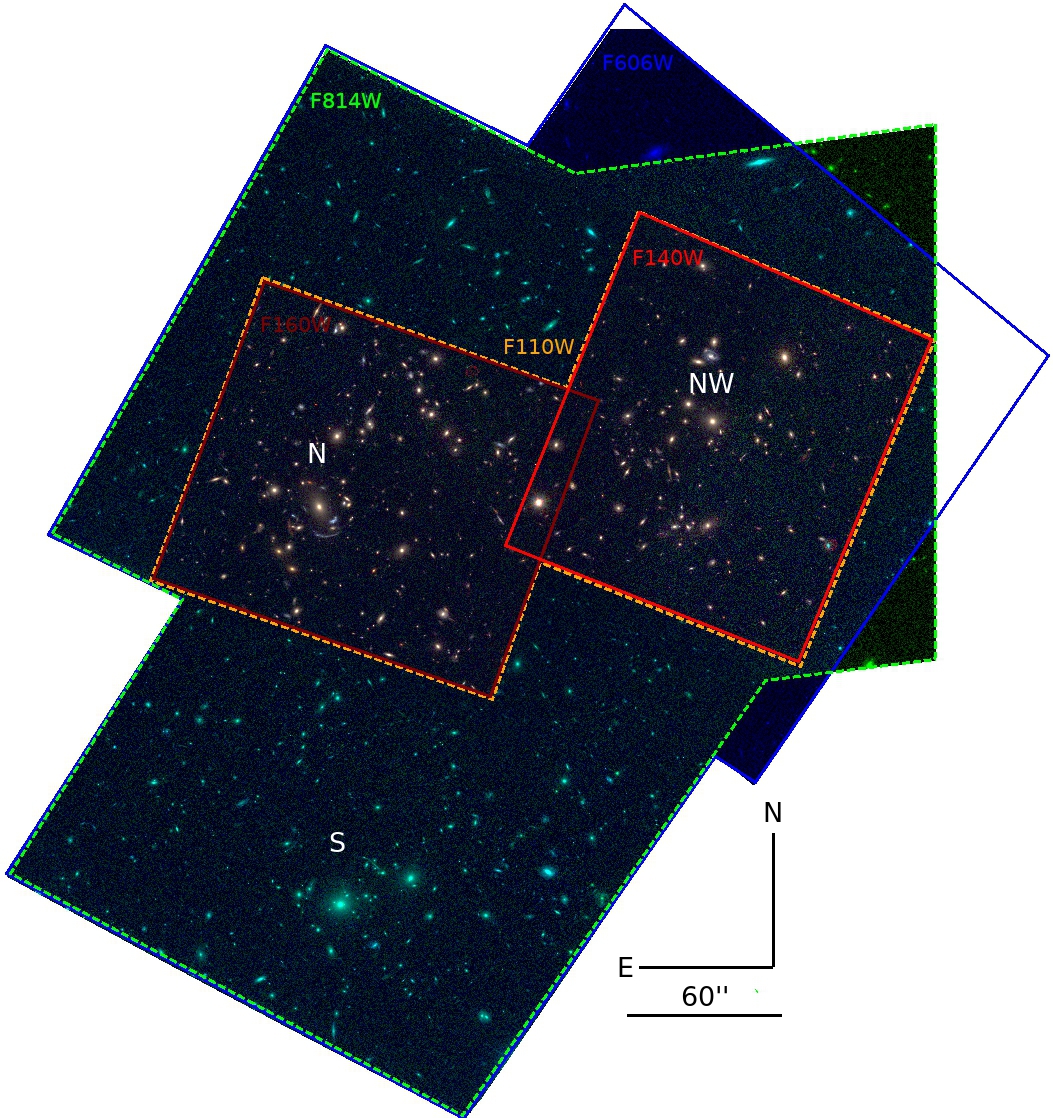}
\caption{The HST observation footprint, including the north cluster \clusternameN, the south cluster, \clusternameS, and the northwest group. The figure is a color composite mosaic of HST WFC3/F110W, ACS/F814W, and ACS/F606W, which provide the largest uniform coverage of these structures. The footprints of the available filters are marked in color. The south cluster is only observed with two filters, while the north and northwest cores have additional WFC3-IR imaging.  
} \label{fig:mosaic}
\end{figure*}

\section{Data}
We use archival HST imaging of \clustername\  
obtained by HST Cycles 18 programs GO-12368 (PI: Morris), GO-SNAP-12166 (PI: Ebeling) and Cycle~25 program GO-15378 (PI: Bayliss).  The datasets and the data reduction procedures are detailed in \citet{Tejos2021}; we provide a short summary here.

We combined these data to create a multi-band mosaic of the field, which contains three structures, hereafter referred to as the north cluster core (\clusternameN), the south cluster core (\clusternameS) and the northwest core.
All three cluster cores are fully covered with ACS/F814 and ACS/F616W, while only the north and northwest cores were observed with WFC3/IR, with F110W coverage of both cores, F160W of only the north core, and F140W of only the northwest core. 
The south core did not get observed with a third filter.  
The data obtained in Cycle~16 (GO-11103; PI: Ebeling) have relatively low signal-to-noise (1200 s total with WFPC2 in F606W) in a region within the ACS footprint, and therefore were not co-added to the ACS/F606W data.

We drizzled and combined all the suitable data onto the same pixel grid to create multi-band mosaics of the cluster field. 
The data reduction followed the standard drizzle process using Drizzlepac\footnote{\url{http://www.stsci.edu/scientific-community/software/drizzlepac.html}}. 
We used  \texttt{astrodrizzle} to process the exposures of each visit, using a Gaussian kernel with a drop size \texttt{final\_pixfrac}=$0.8$. We aligned images from different visits onto a common world coordinate system (WCS) using \texttt{tweakreg}, and applied the WCS solution to all the raw data using \texttt{tweakback} before re-drizzling the frames to form the final mosaic. The final reduced images span the footprint of the combined ACS/F814W pointings, with North up, a pixel scale of $0\farcs03$ per pixel, and WCS matched to the native ACS/F814W data. 

\begin{figure*}
\epsscale{0.55}
\plotone{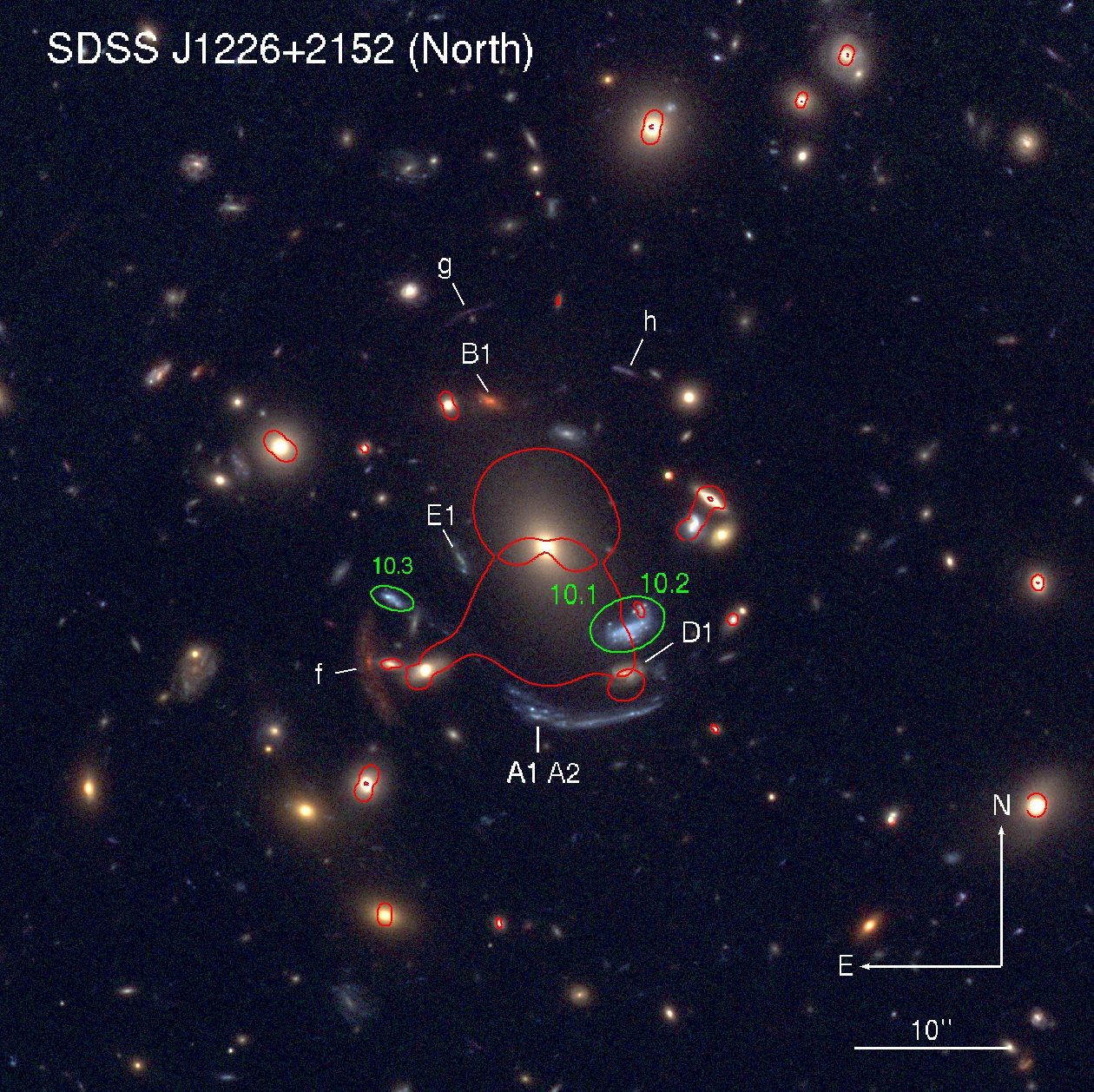}
\plotone{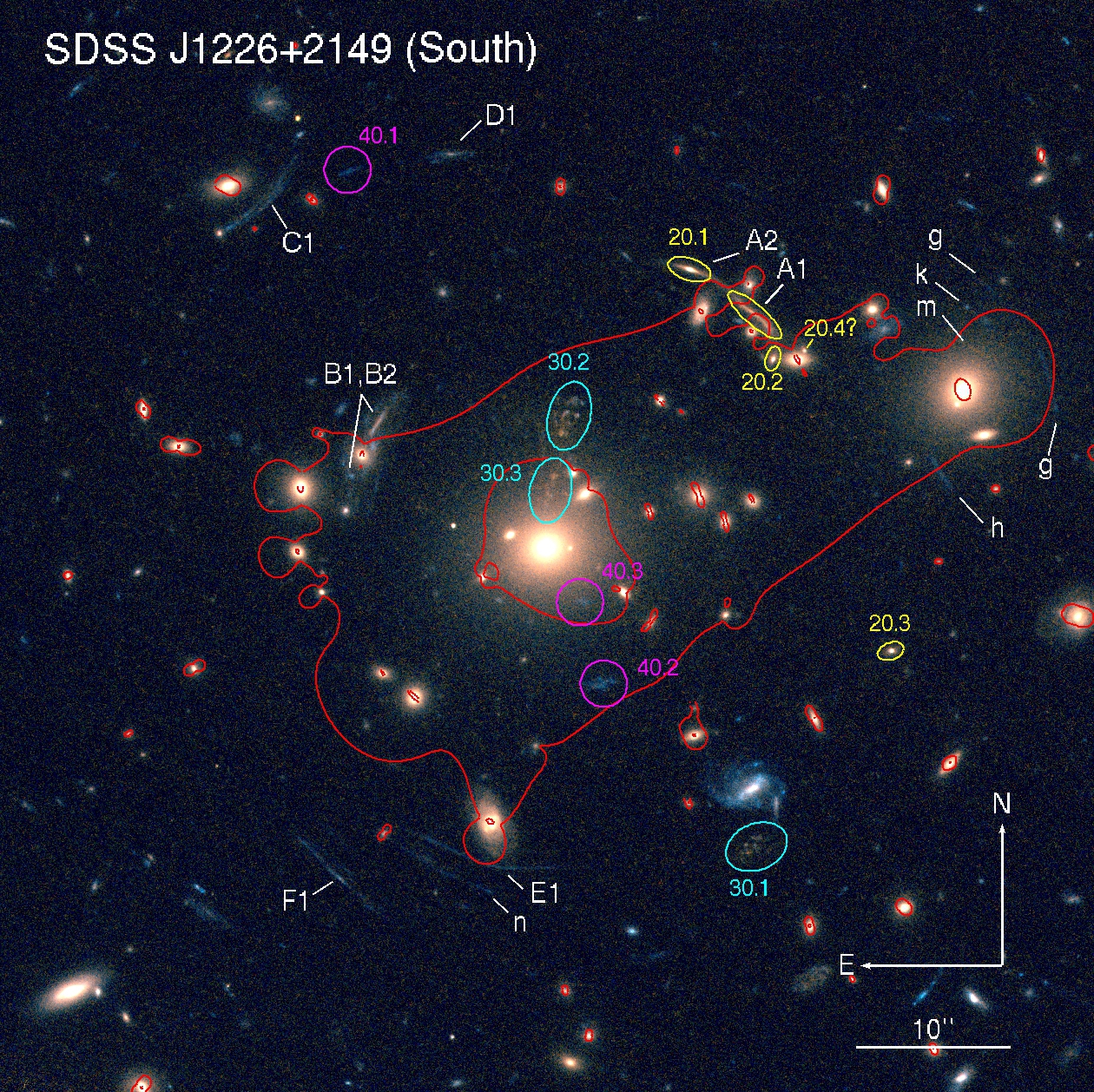}
\caption{\textit{Left:} The north lensing core, \clusternameN, color composite from HST WFC3/F110W, ACS/F814W, and ACS/F606W. \textit{Right:} The south lensing core, \clusternameS, in F606W and F814W only. 
Multiply-imaged lensed galaxies that are used in our analysis are labeled with ellipses and color-coded. Lensed features that were identified by \citet{Bayliss2011} are labeled with upper case letters, and new candidates in lower case letters. The critical curves are overplotted in red for a source at $z=\zarc$ in the north cluster and $z=1.6045$ in the south cluster. } \label{fig:cores}
\end{figure*}

The full field is presented in \autoref{fig:mosaic}. \autoref{fig:cores} shows a color rendition of the north cluster, from F110W (red), F814W (green), and F606W (blue), in the left panel; the south cluster is shown in the right panel, where an extrapolation of F814W and F606W is used in the red channel instead of the missing WFC3-IR.

\section{Selection of Cluster-Member Galaxies}\label{sec:galaxies}
The HST imaging footprint covers the two lensing cores  in F606W and F814W. These two filters provide good color-based selection of cluster-member galaxies via the red-sequence technique \citep{gladdersyee2000}. The F814W band samples the spectral energy distribution redward of the $4000$\AA\ break, and thus it adequately represents the stellar mass.  
We constructed a photometric catalog of the field using Source Extractor \citep{Bertin1996}, in dual-image mode, using the F814W image as the detection image and measuring the \texttt{MAG\_AUTO} magnitudes in both filters within the F814W-selected apertures. We used the following parameters: \texttt{DETECT\_MINAREA}=5 px, \texttt{DETECT\_THRESH}=5 sigma, and \texttt{DEBLEND\_MINCONT}=0.001. We flagged stars and other detector artifacts and removed them from the catalog based on their locus in the \texttt{MU\_MAX} vs \texttt{MAG\_AUTO} plane. 

The HST coordinates of the photometry catalog were then cross-matched with the spectroscopic redshift catalog from \citet{Bayliss2014}, and the nearest object within $0\farcs75$ was selected. We allowed this tolerance in order to account for slight differences in the astrometric solutions and centering between these datasets. We visually inspected the catalog to ensure that no false matches were made. This resulted in 80 galaxies with a spectroscopic redshift within the HST footprint. 

\begin{figure}
\epsscale{1.3}
\plotone{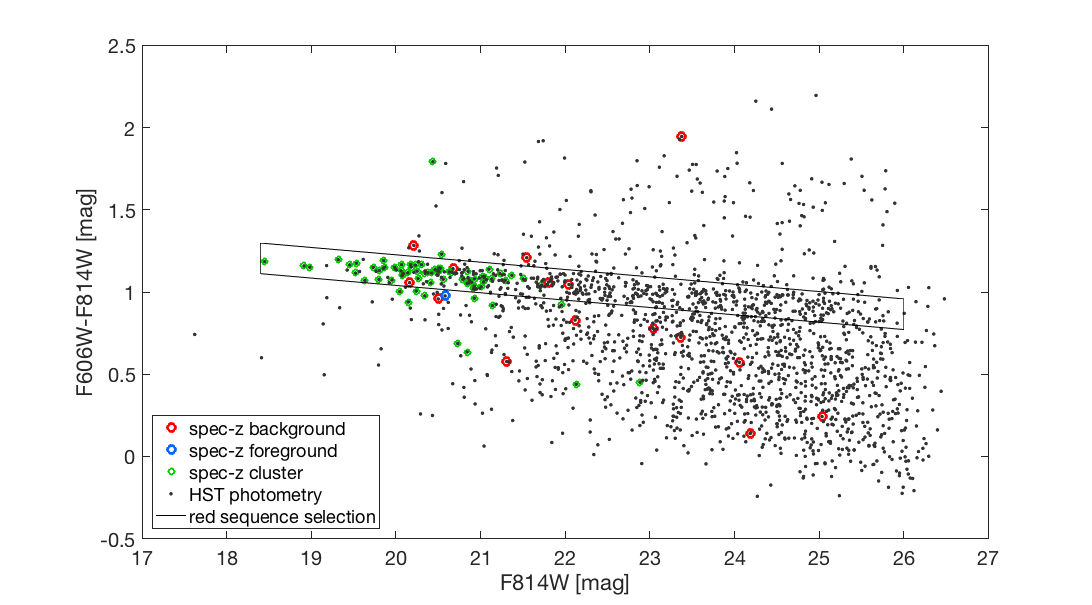}
\caption{Color-magnitude diagram based on HST photometry. The F814W-F606W color of non-stellar objects in the HST footprint is plotted against the F814W magnitude. Galaxies with spectroscopic redshift from \citet{Bayliss2014} and \citet{Bayliss2011} are plotted in green (cluster members), blue (foreground) and red (background). The spectroscopic cluster members were linearly fit with an iterative $3-\sigma$ clipping and the resulting red sequence selection box is shown in black. Apparent red-sequence galaxies with foreground or background spectroscopic redshift were rejected, and bluer or redder galaxies with spectroscopic redshift consistent with the cluster were added back into the cluster-member catalog.  } \label{fig:cmd}
\end{figure}

We construct a F606W--F814W vs F814W color-magnitude diagram of spectroscopic galaxies with $\zspec = \zcluster \pm 0.021$ and fit the cluster red sequence \citep{gladdersyee2000} with a linear fit, using iterative 3-$\sigma$ clipping, in \autoref{fig:cmd}. This iterative process eliminates the blue cluster-member galaxies, and defines the spectroscopically-confirmed red sequence and its width in color space.  
Finally, we applied this color-magnitude selection to the full photometric catalog. We set the faint-end limit at 26 magnitudes in the F814W band.

To account for blue or dusty cluster member galaxies that fall off the red sequence, we supplemented the color-selected catalog with the remaining spectroscopically confirmed cluster member galaxies within the HST footprint.

The galaxy catalog was then cross matched again with the spectroscopic catalog and with the positions of identified lensed features, to eliminate objects with redshifts in the foreground or background. Finally, we manually inspected the catalog for artifacts that were not eliminated in the previous steps, and other objects that are obviously not cluster galaxies such as overly-deblended emission regions in foreground galaxies.
\autoref{fig:cmd} shows the F606W--F814W vs F814W color-magnitude diagram for galaxies in the HST footprint, and the selection of red-sequence cluster members.

\section{Lensing Analysis}
\subsection{Methodology}
The lens plane is complex, with two distinct lensing cores, \clusternameN\ in the north and \clusternameS\ in the south, and a group in the northwest, all within a few hundred km s$^{-1}$ in velocity space \citep{Bayliss2014}. Using ground-based data, \citet{Bayliss2011} identified giant arcs in both SL cores, and obtained spectroscopic redshifts of two strongly-lensed systems and several other background sources. 
The high resolution of the HST data allow us to confirm some of these lensing features as multiply-imaged strongly lensed galaxies, and identify additional strongly lensed galaxies to be used as constraints.

We use the parametric lens modeling algorithm \lenstool\ \citep{jullo07}. This algorithm relies on Markov Chain Monte Carlo formalism to explore the parameter space, and to identify the set of lens-plane parameters that produce the smallest scatter between predicted and observed strong lensing constraints. We model the lens plane with a linear combination of cluster-scale and galaxy-scale projected mass density halos, which are parameterized as pseudo-isothermal ellipsiodal mass distribution \citep[PIEMD, also known as dPIE;][]{eliasdottir07}. The PIEMD profile has seven parameters: position ($x$, $y$), ellipticity $e$, position angle $\theta$, core radius $r_c$, truncation radius $r_{cut}$, and a normalization $\sigma$. The galaxy-scale potentials are fixed to the observed $x$,$y$ coordinates of the cluster-member galaxies, and $e$, $\theta$ are fixed to the properties of the stellar light as measured with Source Extractor (see \autoref{sec:galaxies}). The other parameters are scaled to their F814W luminosity using the scaling relations described in \citet{limousin05}. The parameters of cluster-scale and group-scale halos are usually allowed to vary, except for $r_{cut}$, which for cluster-scale halos is larger than the typical region where lensing evidence is found, and in our case, larger than the distance between the SL cores.

We model the complex lens plane iteratively, solving each SL core separately while fixing the masses of the other main cluster halos. Then, we combine the models and solve for the entire lens plane jointly.

Our starting point is the lens model (hereafter V0) published in \citet{Tejos2021} and used by \citet{Dai2020,Solimano2021}, and \citet{Solimano2022}. That model solved for the lensing potential of the north cluster based on the lensing constraints around that core, with shear from a fixed circular PIEMD halo at the position of the south cluster with normalization $\sigma=1100$ \kms\ (note that $\sigma$ relates to, but is not equal to, the measured velocity dispersion; see \citealt{eliasdottir07}). 
We start by refining the mass halo of the south cluster using the observed lensing constraints around this halo. During this step, we keep the north cluster-scale halo fixed to the best-fit parameters of the V0 model. We add a group-scale mass halo with $\sigma=600$ \kms\ fixed to the position of the northwest cluster. These two fixed halos  generate lensing shear from the directions of the structures with which they are associated; their parameters are only fixed in this initial lens modeling iteration, and will be free in subsequent iterations of the lens model.  We allow all the parameters of the south cluster-scale halo to vary, with the exception of the cut radius, which is fixed to 1500 kpc. The south BCG is decoupled from the other cluster members and its core, cut, and $\sigma$ parameters are left as free parameters. 
We find that the lensing evidence in the south cluster requires an additional mass halo in the vicinity of the third-brightest galaxy in this field, and we therefore free the slope parameters ($r_c$, $r_{cut}$, $\sigma$) of this galaxy as well, while fixing its geometric parameters. 

Once a satisfactory solution is obtained in the south, we allow the parameters of the halos representing the north and northwest clusters to vary as described below. 

The north cluster is dominated by a cluster-scale halo, centered near the BCG. Similarly to the south cluster and the V0 model, we leave all of its parameters free with the exception of $r_{cut}=1500$ kpc. Two galaxies that   appear in close projection to the images of the lensed galaxies are decoupled from the cluster member catalog. 
The first is a faint cluster member galaxy that perturbs  image 10.2 of \arcnamelong\ and generates two additional images of clumps 8 and 9 (see \autoref{sec:arcs} for a description of the lensed images). The second is an interloping galaxy at $z=0.77$ (galaxy G1 in \citealt{Tejos2021}, D1 in \citealt{Bayliss2011}). As was done in \citet{Tejos2021}, we include this galaxy in the same lens plane of the cluster. This approximation simplifies the lensing solution to one lens plane, and improves the accuracy of image positions as well as magnification compared to omitting the line-of-sight mass from the lens model \citep{Raney2020}.  

The final lens model, which we label V1, has seven halos solved for individually: three in the south cluster core (a cluster-scale halo, the BCG, and another luminous galaxy), three in the north cluster core (a cluster-scale halo, a foreground galaxy, and a faint cluster-member galaxy), and one in the northwest cluster core (a group-size halo). The galaxies' PIEMD scaling parameters $\sigma$ and $r_{cut}$ are also allowed to vary.  This model (V1) has $32$ free parameters and $50$ constraints, resulting in image-plane rms of $0\farcs26$. 
For reference, model V0 had 16 free parameters and 26 constraints, and an image-plane rms of $0\farcs08$ for the knots in the lensed galaxy in the north core \citep{Solimano2022}. The best-fit parameters of V1 and their statistical uncertainties are tabulated in \autoref{table.lensmodel}.

\subsection{Strong Lensing Evidence}\label{sec:arcs}
In this section we describe the identification of images of  strongly lensed galaxies in the two SL cluster cores. We label the identified arcs in \autoref{fig:cores}. The coordinates that were used as constraints, and the available spectroscopic redshifts, are tabulated in \autoref{tab:arcstable}. There are hints of arc-like features between the SL cores, but they are either primarily from galaxy-galaxy lensing or otherwise not robust enough to be used as strong lensing constraints.  

The multiple images are labeled with IDs in the form AB.N, where A denotes the source number, B denotes a morphological feature within the source, and N identifies the image number within the set of multiple images of the same source. For example, the ID 15.2 is given to image number 2 of clump 5 in source 1, while 10.2 is assigned to the entire image 2 of source 1. 

\subsubsection{South Cluster}
In the south cluster core, we use SL evidence from three multiply-imaged lensed galaxies. Source~20 was identified and spectroscopically confirmed at $z=1.6045$ by \citet{Bayliss2011}, with the elongated arc and the brightest clump marked as A1 and A2 in their paper, respectively. The high resolution ACS/F814W+F606W imaging reveals that A1 has much lower surface brightness than A2, likely because it is a partial image that does not include the core of the source galaxy. We further identify two counter images with high confidence, marked 20.3 and 20.2, both with similar F616W--F814W color and surface brightness as 20.1 (A2). We use as constraints the core of the galaxy, and regions in the extended arc that match in surface brightness to some of the other images. 
We identify a candidate fourth image (20.4) with similar morphology and color. This image may be a result of contribution to the lensing potential from a nearby cluster-member galaxy. This candidate is not used in the model. 

Source 30 is a low surface brightness clumpy radial arc, composed of two images, with a counter image southwest of the cluster core. The morphology and colors of the three images are consistent with each other and with the expectation from SL geometry. This system was identified by \citet[][labeled A in their paper]{Sulzenauer2021}, who report a redshift from CO lines, $z_{\rm CO}= 1.60454\pm0.00001$, i.e., the same as was measured for source 20. An iteration of the lens model that treated the redshift of this galaxy as a free parameter found it to be consistent with this measurement. We match several emission clumps between the images of the source to be used as constraints.  

Source 40 mirrors the lensing geometry of Source 30, with two radial images in the southwest and a counter image in the northeast. We use the centroids of the F606W emission in each image as constraints, and leave the source redshift as a free parameter.  

Other arc-like features were labeled in \citet{Bayliss2011}, however, we were unable to identify multiple images for them for the purpose of constraining the lens model. Arc B1 is outside the cluster SL regime due to its low redshift ($z=0.8014$), although it does appear to be tangentially distorted and likely locally lensed by a nearby cluster member galaxy. Arcs C1 ($z=0.9134$), D1 ($z=1.1353$), E1, and F1 are likewise tangentially distorted but do not provide useful SL constraints.
We identify a handful of other lensed features in the field, notably around a nearby massive galaxy $\sim29''$ northwest of the BCG, which adds confidence to our inference that this galaxy traces a massive mass component. We mark these arcs as candidates with lowercase letters in \autoref{fig:cores}. 

\begin{figure}
\epsscale{1.1}
\plotone{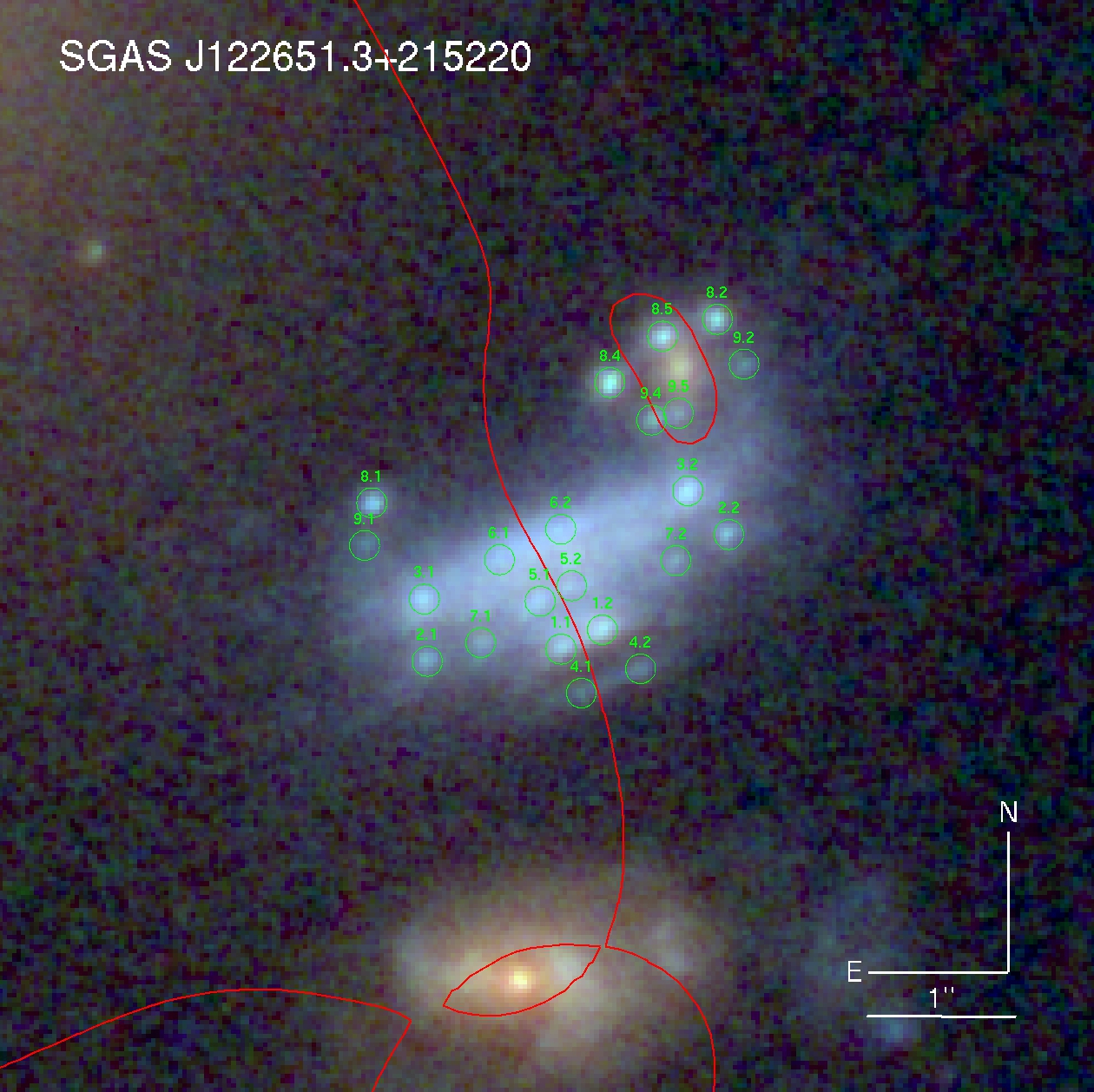}
\caption{Close-up view on the main arc in this field, \arcnamelong, in the north lensing core. The color composite is from HST WFC3/F110W, ACS/F814W, and ACS/F606W. The clumps that were used as lensing constraints are labeled in green. The critical curve, which marks loci of highest magnification for a source at $z=\zarc$, is over-plotted in red. The critical curve defines the axis of symmetry, as multiple images of each clump form on either side of the line bisecting the arc. A cluster member galaxy can be seen near clumps 8 and 9. This galaxy adds complexity to the lensing potential, causing these clumps to form two additional images.  } \label{fig:zoom}
\end{figure}

\subsubsection{North Cluster}
The north cluster core has two prominent blue arcs. Source~10 is a clumpy Lyman break galaxy at $z=$\zarc\ \citep{Koester2010}. It has an extremely bright image formed as a merging pair on opposite sides of the critical curve, southwest of the BCG. The HST imaging reveals two clumpy partial images with striking mirror symmetry, 10.1 and 10.2 in \autoref{fig:cores}. A close view of the merging pair is shown in \autoref{fig:zoom}. A third image of this source, 10.3, appears $16''$ due east of the bright arc. Our lensing analysis indicates that only a small portion at the edge of the source galaxy is mapped to the locations of the bright image (the 10.1,10.2 pair), while 10.3 is a complete image.
This system provides all of the lensing constraints in the north core. We identify nine unique clumps and map them between images 10.1 and 10.2 (see \autoref{fig:zoom}). We further require that the critical curve passes through the point of symmetry, thus adding another lensing constraint at this location. The lower magnification of 10.3 means that the small structures identified in 10.1/10.2 may not be resolved in 10.3 \citep[e.g.,][]{Meng20}, so it is difficult to determine their exact mapping to 10.3. Lensing geometry dictates that if the source is bisected by the lensing caustics in the source plane, it would be the west part of 10.3 that enters the high-multiplicity region while the east part (which also happens to include the brighter core of the galaxy) remains outside of the caustic, forming just one image.
We therefore use as constraint a clump in the west-most edge of image 10.3, that has similar surface brightness as some clumps in images 10.1/10.2. 
Finally, we identify an increased multiplicity of clumps 8 and 9 in image 10.2, which forms due to contribution from a cluster member galaxy that is superimposed on that image. 

A giant arc, labeled A1A2 in \autoref{fig:cores} appears south of the bright images of Source 10. It was spectroscopically confirmed as residing at the same source plane as source~10, $z=2.9233$ \citep{Koester2010,Bayliss2011}; the interaction between these galaxies, and their extended \lya\ halo, were recently studied by \citet{Solimano2022}.  

The ground-based data used by \citet{Koester2010} lack the resolution needed to fully interpret the lensing evidence in this field. Nevertheless, they were able to deduce that the critical curve must pass through the bright arc in order for the lens to not produce counter images that were not observed.  They correctly predicted that space-based imaging will confirm the merging-pair nature of this lensed image. As for the giant arc, it was misinterpreted as two multiple images bisected by the critical curve, with a counter image at 10.3. The HST data reveal that the arc lacks the symmetry that would be expected from such lensing configuration. The giant arc is rather a highly flexured, tangentially distorted \textit{single} image of its source galaxy. As such, it does not provide multiple-image constraints. Nevertheless, the fact that it is singly-imaged can be used to reject models that predict multiple images of the giant arc.

A few other arc-like features appear around this cluster core. An extended red galaxy, with high flexure, can be seen south of 10.3. This image is likely a single image of the source, with no detection of counter images; it is therefore not used as a strong lensing constraint. 
\citet{Bayliss2011} measured the redshifts of three other background sources but their redshifts are too low for them to be strongly lensed (B1 at $z=1.34$, D1 at $z=0.77$, and E1 at $z=0.73$). \citet{Tejos2021} inspected the field of view of the available MUSE data in this field (ESO programme IDs 0101.A-0364(A) and 0102.A-0718(B)), both for galaxies detected by HST and for emission-line-only galaxies with no HST counterpart, but no new constraints were found.

That only one strongly lensed galaxy can be used to constrain this cluster core, albeit with numerous constraints from emission clumps, is a limiting factor.   
The anticipated JWST/NIRCAM and MIRI multiband imaging of this field, as part of TEMPLATES (program number 1355, PI: Rigby) may reveal additional lensed galaxies that are invisible to HST. Such detections can further constrain this cluster core, and reduce the lens modeling uncertainties.

\section{Source Plane Analysis} 
The lensed galaxy at $z=$\zarc\ in the north cluster core is one of the four targets to be studied by \TEMPLATES, and therefore will be of high interest to the community. In this section, we describe a qualitative source plane analysis of this galaxy and its companion, with the goal of describing the mapping between the image plane and the source plane. 

As mentioned in previous sections, the bright arc \arcnamelong\ is  formed by a ``merging pair'' lensing configuration where two partial images of the source appear in close proximity on opposite sides of the critical curve. For a clumpy galaxy like this one, the result is an almost perfectly mirrored double image, with the symmetry axis being the critical curve that bisects it.
The critical curve represents the theoretical loci in the image plane where an infinitely small source will experience an infinitely high magnification. In close proximity to the critical curve, the magnification is extremely high, which leads to the highly resolved image of an intrinsically small physical extent of the source galaxy. The projection of the critical curve to the source plane is referred to as the source-plane caustic. When an extended galaxy is bisected by a caustic in the source plane, regions in the galaxy that fall interior the caustic will have higher multiplicity; every ``caustic crossing'' adds two lensed images that emerge on opposite sides of the corresponding critical curve.

\begin{figure}
\epsscale{1.2}
\plotone{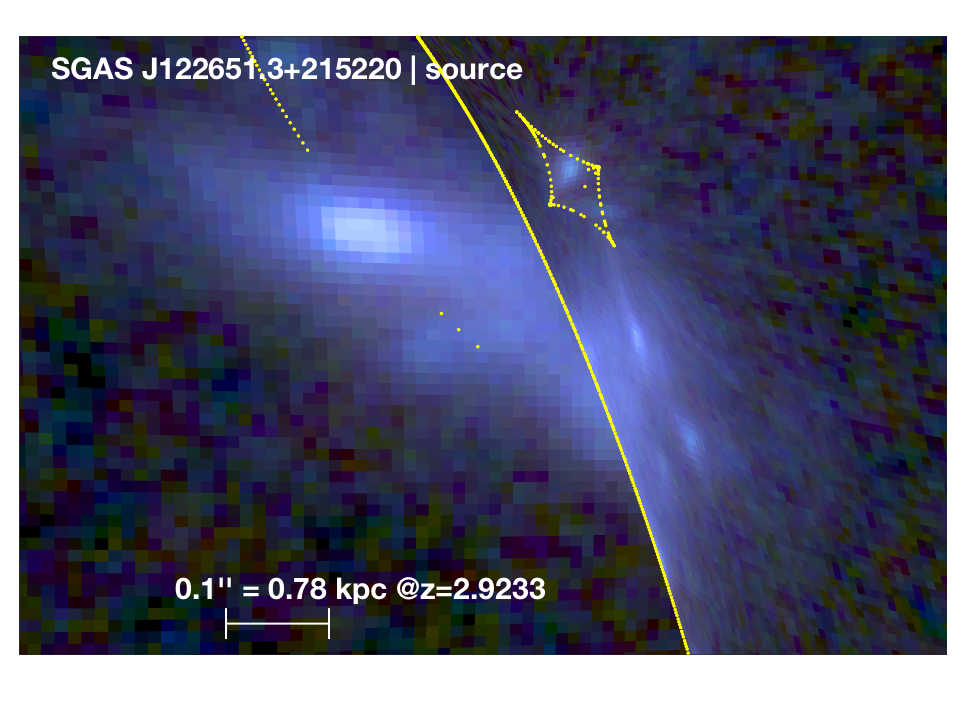}
\caption{Source plane projection of \arcnamelong, obtained by ray-tracing the pixels of the color-rendition image from the image plane to the source plane using the best-fit lens model. The source plane caustic is over-plotted in yellow, marking loci of highest magnification. The caustic bisects the source galaxy; only regions interior to a caustic are multiply imaged: the part of the galaxy left of the main caustic is singly-imaged (10.3), while the parts to the right of the main caustic form three images (10.1, 10.2, 10.3). The clumps that are enclosed within the secondary caustic have a total of five multiple images each.} \label{fig:source}
\end{figure}

\begin{figure}
\epsscale{1.2}
\plotone{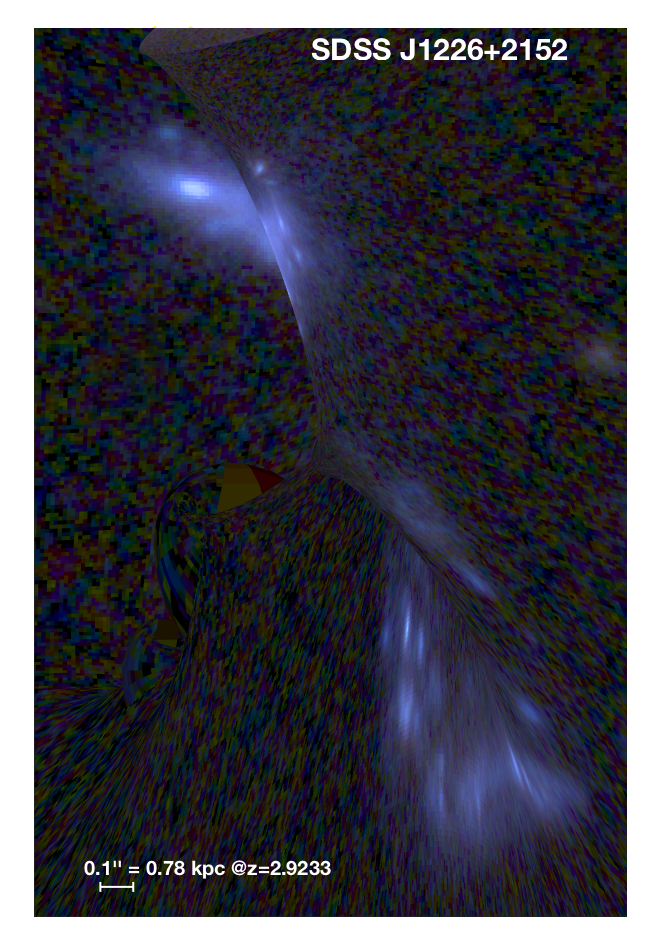}
\caption{Same as \autoref{fig:source}, but zoomed out to show the two interacting galaxies in the source plane. Foreground galaxies were artificially masked out from the color rendition prior to ray-tracing. The caustics are not shown.} \label{fig:sourcemerger}
\end{figure}

In \autoref{fig:source} and \autoref{fig:sourcemerger} we show a qualitative sourceplane projection. This source projection is obtained by ray-tracing the image plane pixels through the best-fit lens model, using the lens equation:
\begin{equation}
    \vec{\beta}=\vec{\theta}-\frac{d_{ls}}{d_{s}}\vec{\alpha}(\vec{\theta}),
\end{equation}
where $\vec{\beta}$ is the $x,y$ location in the source plane; $\vec{\theta}$ is the observed image-plane coordinate; $d_{ls}$ and $d_{s}$ are the angular diameter distance from the lens to the source and from the observer to the source, respectively; and $\vec{\alpha}(\vec{\theta})$ is the lensing deflection at position $\vec{\theta}$.
The critical curves were similarly ray-traced to the source plane, and the resulting caustics are shown in yellow. The partial images 10.1 and 10.2 are traced to the right of the main caustic, and overlayed on the complete image 10.3. The small curved-rhombus caustic that can be seen in \autoref{fig:source} is caused by the lensing potential of a small galaxy near image 10.2. Clumps 8,9 are enclosed by this caustic, resulting in two extra images for each of these clumps. Due to the high magnification close to the caustics, the pixels containing clumps 1,4,5,6 are extremely compressed in the direction perpendicular to the caustic, and are not resolved in this source plane rendition.  

\begin{figure}
\epsscale{1.2}
\plotone{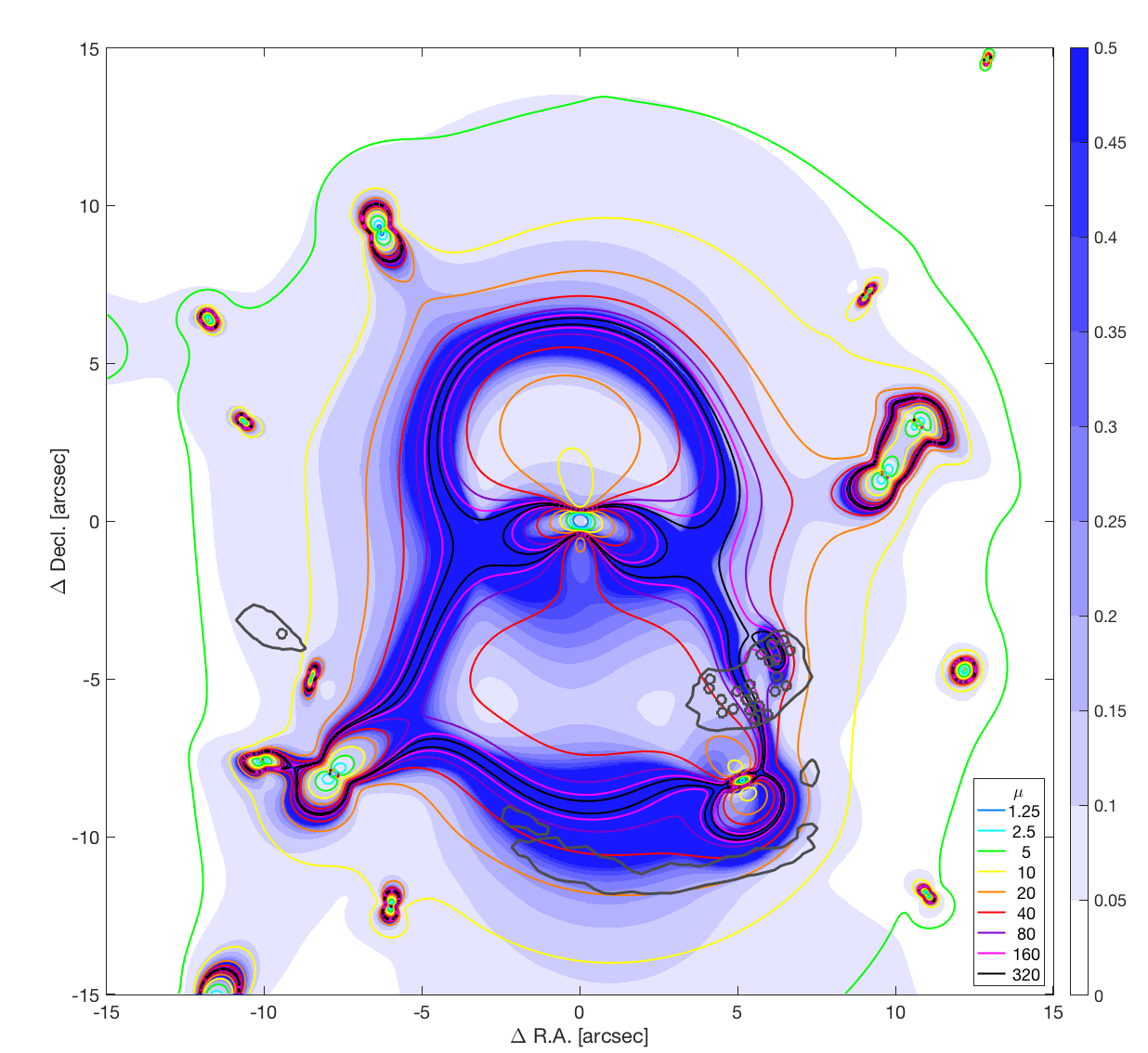}
\caption{The magnification contours for a source redshift $z=\zarc$. The shaded colormap represents the fractional uncertainty, $\sigma$, which is estimated from steps in the MCMC chain, indicating where 68\% of the results fall within $\mu\pm\sigma\mu$. The gray contours mark the location of the lensed galaxies, with gray circles marking the emission clumps, to guide the eye. The image coordinates are measured from R.A.=186.715410, Decl.=21.873718.} \label{fig:magnification}
\end{figure}

\section{Strong Lens Model Deliverables}
This paper accompanies the pre-launch data and data product delivery of the JWST-TEMPLATES program. We provide the community with the best available lens models and lens model outputs; the deliverables described below are available for direct download through MAST via \dataset[10.17909/zqax-2y86]{\doi{10.17909/zqax-2y86}}\footnote{\url{https://archive.stsci.edu/hlsp/templates/}}.
Two versions of the lens model are included with the pre-launch data release: V0 is our original lens model, which was used in \citet{Tejos2021,Dai2020,Solimano2021}, and most recently in \citet{Solimano2022}.  V1 is the improved lens model that is described in detail in this paper. The main difference between these versions is the treatment of the south and northwest cluster cores. In V0 we included the south cluster as a circular mass, fixed to the position of the south BCG, and normalized to the mass estimate from galaxy velocity dispersion. The northwest core was not included. In V1, we model the mass distribution of the south cluster based on lensing constraints from three galaxies, two of them with spectroscopic redshifts. We additionally add the northwest halo, which does not show strong lensing evidence, as contributing external shear. It is included as a circular mass fixed to the position of the northwest BCG, and we leave its normalization as a free parameter.

For each model we provide magnification maps ($\mu$) for a source redshift $z=\zarc$; deflection maps ($\alpha$) in the $x$ and $y$ directions; convergence maps ($\kappa$); and shear maps ($\gamma$). With the exception of the magnification maps, the lensing outputs are linear with the distance term $d_{ls}/d_{s}$ (the ratio of angular diameter distances from the lens to the source and from the observer to the source) and can be scaled to any arbitrary redshift by multiplying $\kappa$, $\gamma$, and $\alpha$ by the desired $d_{ls}/d_{s}$ and dividing by $d_{ls}/d_{s}$ for which it was computed. For $z_l=\zcluster$ and $z_s=\zarc$, $d_{ls}/d_{s}=0.733$.
The magnification can be calculated from $\kappa$ and $\gamma$ for any source redshift, as folllows: 
\begin{equation}
    \mu=\frac{1}{|(1-\kappa)^2-\gamma^2|}.
\end{equation}

In addition to the best-fit maps, we provide $\sim100$ ``range'' files for each lensing output, calculated from $\sim100$ parameter sets drawn from the MCMC sampling of the parameter space. 
These ``range'' files can be used to determine the statistical lens modeling uncertainties of measured properties. \autoref{fig:magnification} shows the magnification in the north core, \clusternameN, for a source at $z=\zarc$, and its uncertainties. The magnification is indicated with contours, and the shaded colormap in the background maps the fractional uncertainty on the magnification. To generate the uncertainty map, for each pixel, we find the range of magnifications spanned by 68\% of the models, and divide this range by the magnification in that pixel from the best-fit model to obtain the fractional uncertainty.
The magnifications of the individual clumps, and their uncertainties, are listed in \autoref{tab:magnification}.

\begin{deluxetable}{lll} 
\tablecolumns{3} 
\tablecaption{Magnification \label{tab:magnification}} 
\tablehead{\colhead{ID} &
            \colhead{$\mu$ } &          
            \colhead{1$\sigma$ uncertainty on $\mu$}       
            \\[-8pt]
            \colhead{} &
            \colhead{} &
            \colhead{[lower upper]}             }
\startdata 
1.1 &  216 & [106 -- 316] \\
1.2 &  206 & [165 -- ...] \\
2.1 &   34 & [25 -- 46] \\
2.2 &   44 & [31 -- 92] \\
3.1 &   45 & [36 -- 60] \\
3.2 &  101 & [45 -- ...] \\
4.1 &  234 & [78 -- 578] \\
4.2 &   87 & [76 -- 321] \\
5.1 &  261 & [175 -- 343] \\
5.2 &  413 & [277 -- 550] \\
6.1 &  156 & [124 -- 277] \\
6.2 &  201 & [116 -- 306] \\
7.1 &   53 & [37 -- 70] \\
7.2 &   61 & [43 -- 122] \\
8.1 &   48 & [43 -- 66] \\
8.2 &   69 & [33 -- 90] \\
8.4 &  168 & [49 -- 256] \\
8.5 &  122 & [22 -- ...] \\
9.1 &   40 & [35 -- 53] \\
9.2 &   49 & [25 -- 66] \\
9.4 &  697 & [60 -- ...] \\
9.5 &  142 & [22 -- ...] \\
\enddata 
\tablecomments{List of magnifications of the clumps in the bright arc. The second column lists the best-fit model-predicted magnifications for a point source located at the exact position of each clump. The brackets indicate the lower and upper magnification corresponding to 1$\sigma$ confidence interval in the parameter space, sampled from the MCMC chain. The magnification for a point source at the counter-image position is 8 [7.2 -- 9.6]. }
\end{deluxetable}

We caution that the statistical uncertainties underestimate the true uncertainties related to lens modeling. Systematic uncertainties are generally unaccounted for by the MCMC sampling process, and are related to different factors that vary from field to field, such as availability and distribution of constraints, spectroscopic redshifts of lensed sources, correlated or uncorrelated structure along the line of sight, and modeling choices
\citep[e.g.,][]{Bayliss2014,rodney15,zitrin15,johnson16,priwe17,meneghetti17,mahler18,remolina18,kelly18, Raney2020}.
The different properties derived by lens modeling are not equally sensitive to statistical and systematic uncertainties. While the mass interior to  the strong lensing evidence (e.g., the mass enclosed by Einstein Radius) is quite robust to lens modeling choices \citep[e.g.,][]{remolina21}, the magnification and time delay can be highly affected, especially in regions close to the critical curve where the magnification gradient is steep. It is recommended to exercise caution in measuring source properties from images near the critical curve, as resolution and sub-pixel variations can be important. 

\vspace{1cm}

\section{Summary and Future Work}
In preparation for the JWST-ERS program \TEMPLATES: \textit{\templateslong} (Program number 1355, PI: Rigby), we present an updated pre-launch lens model for the first science target observed by JWST, the strongly lensed galaxy \arcnamelong. This paper accompanies the TEMPLATES pre-launch data product release. We make available to the scientific community the lens modeling outputs of two versions of this model, an early version of the model, V0, and an updated model, V1,  as high level science products (hlsp) on MAST. These models are based on archival HST imaging, and published spectroscopic redshifts. 

The upcoming and highly anticipated JWST observations of this field will undoubtedly reveal new information on the source, its lensed images, and the entire field. We expect better clump-based analysis and identification to be made feasible, new background sources that are invisible to HST, and yet-unobserved foreground sources.
This new information will be used to update and improve the lens model. Any additional lensed system in the north core, \clusternameN, will greatly benefit the lens model which is currently only constrained by one galaxy at one source redshift. In particular, lensed galaxies north of the BCG, or images buried in the BCG light, will add valuable constraints. As part of TEMPLATES, we will release a JWST-based lens model and lens model outputs to benefit the scientific community and support a host of research avenues that will be made possible by the exquisite combination of this powerful cosmic telescope and JWST. 

\acknowledgements
Based on observations made with the NASA/ESA \HSTlong, obtained at the Space Telescope Science  Institute, which is operated by the Association of Universities for Research in Astronomy, Inc., under NASA contract NAS 5-26555. These observations are associated with programs GO-12166, GO-12368, and GO-15378.
This preparatory work for JWST-ERS program 01355 was funded through a grant from the STScI under NASA contract NAS5-03127.
Support for HST Program GO-15378 was provided through a grant from the STScI under NASA contract NAS5-26555.
This work used the MATLAB Astronomy and Astrophysics Toolbox \citep[MAAT][]{Ofek2014}

%

\facilities{HST(ACS,WFC3)}


\software{Drizzlepac\footnote{\url{http://www.stsci.edu/scientific- community/software/drizzlepac.html}}, Source Extractor \citep{Bertin1996}, \lenstool\ \citep{jullo07}, MAAT \citep{Ofek2014}         }




\vspace{5cm}
\bibliographystyle{yahapj}
\bibliography{bib}

\appendix
\section{Model Constraints}
In \autoref{tab:arcstable} we provide the lensing constraints that were used in the lens model described here, V1. \autoref{table.lensmodel} lists the best-fit model parameters and their statistical uncertainties. 
\startlongtable
\begin{deluxetable*}{llllll} 
\tablecolumns{7} 
\tablecaption{List of lensing constraints and parameters \label{tab:arcstable}} 
\tablehead{\colhead{ID} &
            \colhead{R.A. [deg]}    & 
            \colhead{Decl. [deg]}    & 
            \colhead{$z_{spec}$}     & 
            \colhead{$z_{model}$}       & 
            \colhead{Notes}       \\[-8pt]
            \colhead{} &
            \colhead{J2000}     & 
            \colhead{J2000}    & 
            \colhead{}       & 
            \colhead{}       & 
            \colhead{}             }
\startdata 
\multicolumn{6}{l}{North cluster}\\
\hline
11.1 & 186.713800 & 21.872048 & 2.9233 & \nodata& \\
1.2 & 186.713717 & 21.872084 &  & & \\ 
2.1 & 186.714068 & 21.872025 &  & & \\ 
2.2 & 186.713463 & 21.872261 &  & & \\ 
3.1 & 186.714074 & 21.872141 &  & & \\ 
3.2 & 186.713546 & 21.872343 &  & & \\ 
3.3 & 186.718245 & 21.872722 &  & & \\ 
4.1 & 186.713759 & 21.871966 &  & & \\ 
4.2 & 186.713640 & 21.872011 &  & & \\ 
5.1 & 186.713842 & 21.872137 &  & & \\ 
5.2 & 186.713778 & 21.872166 &  & & \\ 
6.1 & 186.713923 & 21.872214 &  & & \\ 
6.2 & 186.713800 & 21.872271 &  & & \\ 
7.1 & 186.713961 & 21.872060 &  & & \\ 
7.2 & 186.713569 & 21.872213 &  & & \\ 
8.1 & 186.714179 & 21.872320 &  & & \\ 
8.2 & 186.713486 & 21.872662 &  & & \\ 
8.4 & 186.713702 & 21.872544 &  & & \\ 
8.5 & 186.713596 & 21.872630 &  & & \\ 
9.1 & 186.714193 & 21.872241 &  & & \\ 
9.2 & 186.713433 & 21.872579 &  & & \\ 
9.4 & 186.713618 & 21.872474 &  & & \\ 
9.5 & 186.713564 & 21.872487 &  & & \\ 
\hline
\multicolumn{6}{l}{South cluster}\\
\hline
20.1 & 186.710230 & 21.836167 & 1.6045 &\nodata & A2\\
20.2 & 186.708627 & 21.834542 &  & & \\
20.3 &  186.706348 & 21.829278 &  & & \\
c20.4 & 186.708021 & 21.834697  &  & & Candidate\\
22.1 & 186.710050 & 21.836095 &  & & \\
22.2 & 186.708763 & 21.835144 &  & & A1\\
22.3 & 186.709056 & 21.835403 &  & & A1\\
23.1 & 186.710431 & 21.836224 &  & & \\
23.2 & 186.708662 & 21.834474 &  & & \\
\hline
30.2 & 186.712640 & 21.833603 &  1.6045 & 1.65 & \\
30.3 & 186.713029 & 21.832133 &  & & \\ 
31.1 & 186.709164 & 21.825643 &  & & \\ 
31.2 & 186.712747 & 21.833211 &  & & \\ 
31.3 & 186.712906 & 21.832421 &  & & \\
\hline
40.1 & 186.716839 & 21.837948 & \nodata & & \\ 
40.2 & 186.712018 & 21.828723 &  & & \\ 
40.3 & 186.712329 & 21.830201 &  & & \\

\enddata 
\tablecomments{The IDs, positions, and redshifts of clumps within strongly lensed multiply-imaged galaxies that were used as lens modeling constraints.  
The IDs of images of lensed galaxies are labeled as $AB.X$ or $AB.X$ where $A$ is a number indicating the source ID (or system name);  $B$ is a number indicating the ID of the emission knot within the system; and $X$ is a number indicating the ID of the lensed image within the multiple image family. A prefix $c$ identifies candidates. The spectroscopic redshifts of sources 10 and 20 are from \citet{Bayliss2011}. The spectroscopic redshift of source 30 is from \citet{Sulzenauer2021}. 
}
\end{deluxetable*}.

\newcommand{\PxA}	{$-10.72_{-1.20}^{+1.12}$}
\newcommand{\PyA}	{$-151.76_{-0.73}^{+0.73}$}
\newcommand{\PeA}	{$0.17_{-0.05}^{+0.06}$}
\newcommand{\PthetaA}	{$25.3_{-5.3}^{+1.8}$}
\newcommand{\PrcA}	{$161.1_{-36.9}^{+35.8}$}
\newcommand{\PsigmaA}	{$1339.66_{-139.48}^{+116.63}$}
\newcommand{\PrcB}	{$11.8_{-4.7}^{+4.5}$}
\newcommand{\PsigmaB}	{$386.32_{-44.58}^{+13.06}$}
\newcommand{\PcutB}	{$93.1_{-28.4}^{+54.5}$}
\newcommand{\PrcC}	{$0.48_{-0.43}^{+1.47}$}
\newcommand{\PsigmaC}	{$252.3_{-50.8}^{+110.3}$}
\newcommand{\PcutC}	{$75.8_{-52.4}^{+21.1}$}
\newcommand{\PxD}	{$-0.05_{-0.22}^{+0.54}$}
\newcommand{\PyD}	{$-1.61_{-0.85}^{+0.09}$}
\newcommand{\PeD}	{$0.03_{-0.03}^{+0.01}$}
\newcommand{\PthetaD}	{$36.1_{-5.6}^{+32.5}$}
\newcommand{\PrcD}	{$35.8_{-0.7}^{+3.9}$}
\newcommand{\PsigmaD}	{$623.1_{-13.5}^{+25.6}$}
\newcommand{\PeE}	{$0.36_{-0.16}^{+0.03}$}
\newcommand{\PthetaE}	{$9.5_{-12.0}^{+49.3}$}
\newcommand{\PrcE}	{$0.93_{-0.12}^{+0.07}$}
\newcommand{\PsigmaE}	{$102.5_{-19.5}^{+26.8}$}
\newcommand{\PcutE}	{$74.7_{-37.3}^{+16.8}$}
\newcommand{\PeF}	{$0.28_{-0.15}^{+0.51}$}
\newcommand{\PthetaF}	{$104.7_{-14.1}^{+5.1}$}
\newcommand{\PrcF}	{$1.9_{-1.3}^{+0.1}$}
\newcommand{\PsigmaF}	{$57.0_{-19.3}^{+6.7}$}
\newcommand{\PcutF}	{$6.4_{-1.2}^{+8.2}$}
\newcommand{\PsigmaG}	{$436.5_{-30.4}^{+188.1}$}
\newcommand{\Pz}	{$1.95_{-0.08}^{+0.14}$}
\newcommand{\Pcutgal}	{$30.4_{-9.0}^{+9.2}$}
\newcommand{\Psiggal}	{$139.8_{-8.1}^{+6.7}$}

\begin{deluxetable*}{clccccccc} 
\tablecolumns{9} 
\tablecaption{Best-fit lens model parameters \label{table.lensmodel}} 
\tablehead{\colhead{No. }   & 
            \colhead{Component }   & 
            \colhead{$\Delta$ R.A. ($\arcsec$)}     & 
            \colhead{$\Delta$ Decl. ($\arcsec$)}    & 
            \colhead{$e$}    & 
            \colhead{$\theta$ (deg)}       & 
            \colhead{$r_{\rm core} $ (kpc)} &  
            \colhead{$r_{\rm cut}$ (kpc)}  &  
            \colhead{$\sigma_0$ (km s$^{-1}$)}             } 
\startdata 
1&South: Cluster & \PxA & \PyA & \PeA & \PthetaA & \PrcA & [$1500$] &\PsigmaA\\
2&South: BCG     & [$-7.901054$] & [$-153.1872$] & [$0.169$] & [$13.21$] & \PrcB & \PcutB  &\PsigmaB\\
3&South: galaxy & [$-34.967991$] & [$-143.0852$] & [$0.156$] & [$41.82$] & \PrcC & \PcutC  &\PsigmaC\\
4&North: Cluster & \PxD & \PyD & \PeD & \PthetaD &\PrcD & [$1500$] &\PsigmaD \\ 
5&North: BG galaxy & [$-5.123157$] & [$-8.2350$] & \PeE & \PthetaE & \PrcE  & \PcutE & \PsigmaE\\
6&North: galaxy & [$-6.172000$] & [$-4.1250$] & \PeF & \PthetaF & \PrcF  & \PcutF & \PsigmaF\\
7&Northwest: Group & [$-151.576256$]&[$32.6408$] &[$0$] &[$0$] &[100]& [$1500$] & \PsigmaG \\
 &L* galaxy  & \nodata & \nodata & \nodata & \nodata & \nodata  &     \Pcutgal &  \Psiggal  \\ 
\enddata 
\tablecomments{Coordinates are tabluated in arcseconds from a reference coordinate in the North cluster, [R.A., decl.]=[186.715410, 21.873718].  All the mass components are parameterized as PIEMD (see text), with ellipticity expressed as $e=(a^2-b^2)/(a^2+b^2)$. $\theta$ is measured North of West. Statistical uncertainties are inferred from the MCMC optimization and correspond to a 95\% confidence interval. Parameters in square brackets were not optimized. The
location and the ellipticity of the matter clumps associated with cluster galaxies were kept fixed according to their light distribution, and the other parameters determined through scaling relations (see text). The BG galaxy is assumed to be at the same lens plane as the cluster (see text). The image plane rms of the best-fit model is $0\farcs26$.}
\end{deluxetable*} 

\end{document}